\documentclass[a4paper,11pt]{article}
\usepackage{latexsym}
\begin{document}

\newtheorem{definition}{Definition}
\newtheorem{theorem}{Theorem}
\newtheorem{proposition}{Proposition}
\newtheorem{remark}{Remark}
\newtheorem{corollary}{Corollary}
\newtheorem{lemma}{Lemma}
\newtheorem{observation}{Observation}
\newtheorem{fact}{Fact}
\newtheorem{example}{Example}

\newcommand{\qed}{\hfill$\Box$\medskip}

\makeatletter
\def\adots{\mathinner{\mkern 1mu\raise \p@ \vbox{\kern 7\p@ \hbox{.}}
\mkern 2mu \raise 4\p@ \hbox{.}\mkern 2mu\raise 7\p@ \hbox{.}\mkern 1mu}}
\makeatother
\newcommand{\ntowers}{NSPACE$\left(2^{2^{\adots^{2}}}\right)$
 }

\title{On Practical Regular Expressions\\
       {(Preliminary Report)}} 
\author{Holger Petersen\\
Reinsburgstr.~75\\
70197 Stuttgart\\
Germany} 

\maketitle

\begin{abstract}
We report on simulation, hierarchy, and decidability results for Practical Regular Expressions (PRE), which may include back references in addition to the standard operations union, concatenation, and star.

The following results are obtained:
\begin{itemize}
\item PRE can be simulated by the classical model of nondeterministic finite automata with 
sensing one-way heads. The number of heads depends on the number of different variables in the expressions.
\item A space bound $O(n\log m)$ for matching a text of length $m$ with a PRE with $n$ variables
 based on the previous simulation. This improves the bound $O(nm)$ from (C{\^ a}mpeanu and Santean 2009).
\item PRE cannot be simulated by deterministic finite automata with at most three
 sensing one-way heads or deterministic finite automata with any number of
non-sensing one-way heads.
\item PRE with a bounded number of occurrences of variables in any match can be simulated
 by nondeterministic finite automata with one-way heads.
\item There is a  tight hierarchy of PRE with a growing number of non-nested variables over a fixed alphabet.
 A previously known  hierarchy was based on nested variables and growing alphabets (Larsen 1998).
\item Matching of PRE without star over a single-letter alphabet is NP-complete.
This strengthens the corresponding result for expressions over 
larger alphabets and with star (Aho 1990).
\item Inequivalence of PRE without closure operators is $\Sigma^{P}_2$-complete.
\item The decidability of universality of PRE over a single letter alphabet is linked to the existence of  Fermat Primes.
\item Greibach's Theorem applies to languages characterized by PRE.
\end{itemize}
\end{abstract}

\section{Introduction}
Regular expressions have evolved from a tool for the analysis of nerve nets \cite{Kleene56}
into an important domain-specific language for describing patterns. The original
set of operations (union, concatenation, and closure operator star) of what will be called
``Classical Regular Expressions'' (CRE) here has been enhanced in several 
different ways.
A natural additional operation is complementation, leading to Extended Regular Expressions (ERE).
On the language level complementation does not add power to regular expressions, since
any regular expression can be converted into an equivalent finite finite automaton, which
in turn can be complemented via the 
power-set construction. The matching problem however 
becomes harder \cite{JR91,Petersen00} (under the assumption that the corresponding 
complexity classes are different, see 
the table below) and an equivalence test certainly becomes infeasible (its memory requirement
grows faster than any exponential function, \cite{SM73,Fuerer78}). An operation that leads to a less complex equivalence problem is intersection 
(complete in exponential space, \cite{Hunt73b,Robson79}). Expressions based on the operations
union, concatenation, star, and intersection are called Semi-Extended Regular Expressions (SERE).

Practical Regular Expressions (PRE) including several extensions
have been implemented in operating system commands,
data base query languages, and text editors. PRE include many syntactical enhancements like 
notations for sets of symbols (wildcards, ranges, enumerations of symbols etc.), optional subexpressions
or bounded repetition of subexpressions. 
The latter operation has been investigated for the special case of squaring in \cite{MS72}, where completeness in exponential space was shown.
An extension of PRE that goes beyond regular languages in expressive power is the use of back references. 
The $k$-th subexpression put into parentheses (counting opening parentheses from left to right)
can be referenced by $\backslash k$, 
which matches the same string that is matched by the subexpression.
An expression is invalid if the source string contains less than $k$ subexpressions 
preceding the $\backslash k$.
%http://docs.oracle.com/cd/B12037_01/server.101/b10759/ap_posix001.htm#i690819
This is the notation
employed in many implementations, while the definition of \cite{Aho90} admits variable names. 
Also the exact semantics of the expressions vary. In \cite{Larsen98} several syntactical criteria are imposed on
PRE such that no variable can be used before it is defined (this can happen when a variable 
is defined inside of an alternation or Kleene closure that is not part of the matching).
In contrast \cite{CS09} allows for such a situation and assigns the empty set to the variable. 
We will adopt the latter definition here that will
prevent a match with an uninstantiated variable but does not require a syntactical analysis of
variable assignments. This definition is also found in several implementations of PRE.
  
As an example of a non-regular (and not even context-free) language consider the expression 
$$ ((a|b)*)\backslash 1$$
characterizing the language $L_d = \{ ww \mid w\in\{ a, b\}^*\}$ of  ``double-words''. 
Many regular languages can be defined more succinctly by PRE than by CRE, like
$$ (a|b|\ldots)*(a|b|\ldots)(a|b|\ldots)*\backslash 2 (a|b|\ldots)*$$
which characterizes all strings over $\Sigma = \{ a, b, \ldots\}$ with (at least) one repeated occurrence 
of a symbol. This expression has a size linear in $|\Sigma|$, while a CRE requires
size $\Omega(|\Sigma|^2)$.

% Example 3 from Larsen 1998: (a|b)^*%\alpha .  (a|b)^*%\beta . (\alpha|\beta) seems to generate
% every string (take beta empty, \alpha arbitrary, choose \beta in OR), 
% characterization uuv, uvv does not match the structure of the expression (although also correct).

The increase in expressive power of PRE has an impact on the complexity of decision problems.
A central problem is the Matching-Problem:
\begin{description}
\item[Matching-Problem:] Given a PRE $\alpha$ and a string $s$, does a string in the language
described by $\alpha$ occur as a substring of $s$?
\end{description}
The Matching-Problem easily reduces to membership by tranforming the 
expression into $\Sigma^*\alpha\Sigma^*$, where $\Sigma$ is the underlying alphabet. 
Conversely, membership can be reduced to the Matching-Problem by embedding both 
$\alpha$ and  $s$ into special marker symbols or marker strings not occurring otherwise. 
This technique requires at least a binary alphabet. 
The Matching-Problem is known to be NP-complete
\cite{Aho90} and equivalence is even undecidable \cite{Freydenberger11}.

Some results and references are summarized in the following table:

\begin{center}
\begin{tabular}{|l|c|c|c|}\hline
                                     &  PRE                                                            & ERE  & CRE\\\hline
matching,                     &  NP-complete                          & P-complete &     NL-complete          \\
member                &  \cite[Thm.~6.2]{Aho90}  & \cite[Thm.1]{Petersen00}, &         \cite[Thm.~2.2]{JR91}            \\\hline
equivalence         &   undecidable               & NSPACE$\left(2^{\left.2^{\adots^{2}}\right\}{\scriptsize g(n)}}\right)$  & PSPACE-compl. \\
                                        &      \cite[Thm.~9]{Freydenberger11}    & $g(n) = n$ \cite{SM73} (u.b.)  &   \cite[Lem.~2.3]{MS72}  \\
                                     &                                                              & $g(n) = \frac{c\cdot n}{(\log^*n)^2}$ \cite{Fuerer78} (l.b.)   &   \\\hline
non-                             &      $\in$ALOGTIME  &  &      $\in$ALOGTIME      \\
emptiness                     &             & see equivalence  &         \\
                                     &     (see   CRE )                      &  &     \cite[Intr.]{Petersen00}\\\hline  % in P \cite[Lem. 3.1]{Hunt73b}% Petersen00 p. 24 
\end{tabular}
\end{center}

\section{Simulation Results}
The following simulations will be carried out by finite automata that are equipped with 
several read-only heads that scan the input. If the automata can detect coincidence of heads,
they are called ``sensing''.
\begin{theorem}\label{simulationPREmultheadsensing}
Every PRE with $k$ different variables can be simulated by a nondeterministic finite  
automaton with $2k+2$ sensing one-way heads.
\end{theorem}
{\bf Proof.} 
The simulating NFA $M$ stores the PRE $\alpha$ of length $n$ in its finite control
with one of the $n+1$ positions 
marked (initially the position before the PRE). The heads form pairs 
$\ell_i, r_i$ for $1 \le i \le k+1$, where pair $i \le k$ corresponds to variable $v_i$. 
All heads move in parallel along the 
input string while $M$ parses $\alpha$ moving the marked position
until an opening parenthesis is encountered. In this case head $\ell_1$ remains stationary while
the other heads advance. When the matching closing parenthesis is encountered,
head $r_1$ remains at the current position. In this way the value of $v_1$ is
stored and similarly for each of the  variables a sub-string of the input is marked.
 If  $v_i$ occurs in $\alpha$, $M$ leaves 
head $\ell_{k+1}$ at the current position and advances $\ell_i$ comparing the symbols read with
the input using head $r_{k+1}$. Notice that at least this head
is available for the comparison. 
When 
$\ell_i$ and $r_i$ meet, the value of $v_i$ has been compared to the input and
a copy of $v_i$ is marked by $\ell_{k+1}$ and $r_{k+1}$. Now $M$ advances 
$\ell_i$ to the position of $\ell_{k+1}$ and then $\ell_{k+1}$ and $r_i$ to the 
position of $r_{k+1}$. In this way the value of $v_i$ is again marked. 

If in this way $M$ is able to scan all of its input, it accepts. If a mismatch is detected,
$M$ rejects.
\qed

\begin{theorem}\label{inclusionPREmulthead}
The class of languages accepted by nondeterministic finite automata with 
sensing one-way heads properly includes those characterized by PRE.
\end{theorem}
{\bf Proof.} 
Inclusion follows from Theorem~\ref{simulationPREmultheadsensing}.

The language 
$$S = \{ a^iba^{i+1}ba^k \mid k = i(i + 1)k' \mbox{ for some $k' > 0$, $i > 0$}\}$$
is shown not to be generated by any PRE in \cite{CN09}. A deterministic finite automaton with 
$3$ sensing one-way heads can check divisibility of the length of the trailing block of $a$s by 
$i$ or $i+1$ respectively (move heads with distance $i$ or $i+1$  over the block marking the 
last position with the third head). 
Since languages accepted by these finite automata are closed under intersection, the claim follows. \qed 

Finite multi-head automata can be simulated in nondeterministic logarithmic space. 
We obtain the following improvement of  \cite[Corollary 7]{CS09}:

\begin{corollary}
The uniform membership problem for PRE has nondeterministic space complexity $O(n\log m)$ 
where $n$ is the number of pairs of parentheses in the PRE and $m$ is the length of text.
\end{corollary}

It is natural to ask whether a simpler model of computation than nondeterministic finite automata with 
sensing heads can simulate PRE. The next results provide partial answers.
\begin{theorem}\label{impossiblePREdetmulthead}
Languages characterized by PRE with one variable cannot in general
be accepted by deterministic finite automata with at most three sensing one-way heads
or any number of non-sensing one-way heads.
\end{theorem}
{\bf Proof.} Consider the language 
$$M = \{ p\#t_1pt_2 \mid  p, t_1, t_2\in\{ 0, 1\}^*\}$$
formalizing the string-matching problem of deciding whether pattern $p$ occurs in a 
given text $t$. Notice that answers for more realistic problems like reporting the first 
or even every position of $p$ in $t$ would also solve the membership problem of $M$.

A PRE specifying $M$ is
$$((0|1)*)\#(0|1)*\backslash 1(0|1)*.$$
By the result in \cite{GL94} string-matching cannot be done with three sensing 
heads and by the result in \cite{JL96} string-matching cannot be done 
with any number of non-sensing one-way heads by deterministic finite automata. \qed

The above result shows that even very simple PRE cannot be simulated by deterministic 
finite automata with non-sensing one-way heads. This is not true for 
nondeterministic automata. We first define a notion of complexity of PRE depending on
the number of variables occurring in a match of an expression. Let $c$ be the following function
from PRE to $I\!\! N \cup \{ \omega\}$ (where $\omega$ is greater than any
element of $I\!\! N$):
\begin{eqnarray*}
c(\backslash n) & = & 1 \quad\mbox{ for $n\in I\!\! N$}\\
c(a) & = & 0  \quad \mbox{ for $a\in\Sigma$}\\
c(\alpha|\beta)  & = &  \max(c(\alpha), c(\beta))\\
c(\alpha\beta)  & = &  c(\alpha) + c(\beta)\\
c(\alpha*)  & = &  
\begin{array}{cl}
\omega & \mbox{ if  $c(\alpha) \ge 1$}\\
0  & \mbox{ if  $c(\alpha) = 0$}
\end{array}\\
\end{eqnarray*}

\begin{theorem}\label{simulationPREmulthead}
Every PRE $\alpha$ with $k < \omega $ occurrences of variables
($c(\alpha) = k$)
can be simulated by a 
nondeterministic finite automaton with 
$2k+1$ (non-sensing) one-way heads.
\end{theorem}
{\bf Proof.} One head of the simulator $A$ is used for matching the 
regular expression with the input.
The definition of the $i$-th variable is marked on the input by a pair of heads 
for every occurrence of $\backslash i$. 
When a $\backslash i$ occurs, then the trailing head of the pair 
simulating this occurrence is moved along the input comparing the 
segments of the input until $A$ guesses coincidence of the two heads.
Then $A$ moves both heads in parallel until at least one of them 
reaches the right end-marker. If they do not reach the end-marker
at the same time, then $A$ rejects. \qed

\section{Hierarchy Results}
In \cite{Larsen98} languages defined by PRE with a growing number of variables 
are investigated. The complexity of these languages depends on two features of the
construction: 
\begin{itemize}
\item The cardinality of alphabets increases with the number of variables.
\item Variables are nested.
\end{itemize}

Using the simulation from the proof of Theorem~\ref{inclusionPREmulthead} we can establish:
\begin{theorem}\label{hierarchyPREmulthead}
The class of languages characterized by PRE forms an infinite hierarchy with respect to 
the number of variables, even when restricting
PRE to a fixed alphabet and non-nested variables.
\end{theorem}
{\bf Proof.}
It is immediate from the definition that languages characterized by PRE with $k$ variables
form a subset of those characterized by PRE with $k' \ge k$ variables.

The language
$$L_b = \{ w_1\#w_2\#\cdots \#w_b\#w_b\#\cdots
  w_2\#w_1
\mid \forall 1\le i\le b:
  w_i\in\{ 0, 1\}^*\}$$

can be defined by
$$((0|1)*)\#((0|1)*)\cdots((0|1)*)\#\backslash 2b-1\#\backslash 2b-3\cdots
  \backslash 3\# \backslash 1$$
with $b$ subexpression $((0|1)*)$. In \cite{YR78} it is shown that 
nondeterministic finite automata with $h$ one-way heads cannot accept $L_b$ for 
$b > {h\choose 2}$. The main result of  \cite{YR78} also holds for 
sensing heads, see the remark on p.~337.

Every language characterized by a PRE with $k$ variables
can be accepted by a nondeterministic finite automaton with $h = 2k+2$ 
sensing one-way heads
according to Theorem~\ref{inclusionPREmulthead}.
Let $b = {h\choose 2} + 1 = {{2k+2}\choose 2} + 1$. 
Now $L_b$ is a language characterized by a PRE with $b$ variables that cannot
be characterized  with the help of at most $k$ variables. \qed

We now improve the coarse separation of 
the previous hierarchy result using the concept of Kolmogorov 
complexity. The argument is again based on the languages $L_b$, 
but we will show that $L_b$ requires $b$ variables.

\begin{theorem}\label{hierarchyPREmulthead2}
The class of languages characterized by PRE with $b$ non-nested variables properly includes those
characterized by PRE with $b-1$ variables for every $b> 0$.
\end{theorem}

First we prove the following {\em Non-Matching Lemma} for an input $x \in L_b$ that is based
on an incompressible string $w$, see \cite{LV90} for definitions.
Let $w\in\{ 0, 1\}^*$ be an 
incompressible string of length $n$ sufficient large with $n$ a multiple of $b$ 
resulting in integer valued formulas in the following definitions.
Split $w$ into blocks of equal length $w = w_1\cdots w_b$.
Form a word 
$$x =  w_1\#w_2\#\cdots \#w_b\#w_b\#\cdots w_2\#w_1 \in L_b$$
with $|w_i| = n/b$ for all $1\le i \le b$.

\begin{lemma}[Non-Matching Lemma]
Suppose that variable $v$ is instantiated to a substring of length at least $n/b + 2 + 18\log n$ when
matching string $x \in L_b$ as defined above. Then $v$ cannot match any other substring of $x$.
\end{lemma}
{\bf Proof.}
For a contradiction suppose that $v$ matches another substring of $x$. We will identify
$v$ and the value assigned to it in the following discussion. Since $v$ includes
at least two symbols $\#$, the structure of $v$ is $w'\#w_i\#w''$ for some
$1 \le i \le b$ and $w', w'' \in \{ 0, 1, \#\}^*$. We will show how to compress $w$ by 
copying a substring of $w$ from position $p_1$ through $p_2$ to a position $p_3$ in 
$x$.

The first case we consider is that the portion $\#w_i\#$ of $v$ matches a 
substring $\#w_j\#$ for $j \neq i$
(in the left or right half of $x$). Then we can take $p_1$ and $p_2$ as the positions of the first
and the last symbol of $w_i$ and $p_3$ as the position of the first symbol of $w_j$.

The other case is that $v$ matches the corresponding $w'\#w_i\#w''$ from the other  half of $x$.
By the length condition on $v$, at least one of $|w'|, |w''|$ is not less than $9\log n$. Suppose
$|w'| \ge 9\log n$ and $i = b$. Then $w'$ is a suffix of $w_1\#w_2\#\cdots \#w_b$ as well as 
of $w_1\#w_2\#\cdots \#w_{b-1}$ and we can take $p_1 = n - n/b - 9\log n$, 
$p_2 = n - n/b$, and $p_3 = n - 9\log n$. If $i < b$ then $w'$ is a suffix of 
$w_1\#w_2\#\cdots \#w_{i-1}$ as well as of $w_1\#w_2\#\cdots \#w_b\#w_b\cdots\#w_{i+1}$. 
We take $p_1 = (i-1)n/b - 9\log n$,  $p_2 = (i-1)n/b$, and 
$p_3 = (i+1)n/b - 9\log n$.
Now consider the case $|w''| \ge 9\log n$. If $i = b$ then $w''$ is a 
prefix of $w_b\cdots\#\cdots w_2\#w_1$ and of $w_{b-1}\cdots\#\cdots w_2\#w_1$.
We take $p_1 = (b-2)n/b + 1$, $p_2 = (b-2)n/b + 9\log n$ and 
$p_3 = (b-1)n/b + 1$. If  $i <  b$ then $w''$ is a 
prefix of $w_{i+1}\cdots\#\cdots w_2\#w_1$  as well as of $w_{i-1}\cdots\#\cdots w_2\#w_1$.
We take $p_1 = (i-2)n/b + 1$, $p_2 = (i-2)n/b + 9\log n$ and $p_3 = i n/b + 1 $.

The string $w$ can be reconstructed from the following information:
\begin{itemize}
\item A formalization of this description including the recovery algorithm below ($O(1)$ bits).
\item The values of $n$, $p_1$, $p_2$, and $p_3$ in self-delimiting binary form ($8\log n$ bits).
\item $w_1w_2\cdots w_b$ with the portion of length $p_2 - p_1 + 1$ starting at 
position $p_3$ deleted (at most $n - 9\log n$ bits).
\end{itemize}
The string $w = w_1w_2\cdots w_b$ can be reconstructed by copying $p_2 - p_1 + 1$ symbols
starting at  position $p_1$ inserting them at position $p_3$. For $n$ sufficiently large a compression by $\log n - O(1)$
bits is obtained, contradicting the choice of $w$. \qed

We now continue the proof of Theorem~\ref{hierarchyPREmulthead2}.
Let $\alpha$ be a PRE with at most $b-1$ variables characterizing $L_b$. 
Fix a matching of $\alpha$ and $x$ by recording the corresponding alphabet
symbols of $\alpha$ and $x$. 
The string $w$ can be reconstructed from the following information:
\begin{itemize}
\item A formalization of this description including the recovery algorithm below ($O(1)$ bits).
\item The value of $n$ ($O(\log n)$ bits).
\item PRE $\alpha$ ($O(1)$ bits).
\item The occurrence of $\#$ in $\alpha$ matching the center of $x$ ($O(1)$ bits). 
\item For every variable $v_i$ its state when the center has just been matched:
\begin{enumerate}
\item\label{instlong} $v_i$ is instantiated and $|v_i| \ge n/b + 2 + 18\log n$.
\item\label{instshort} $v_i$ is instantiated and $|v_i| <  n/b + 2 + 18\log n$.
\item\label{partinstlong} $v_i$ is partially instantiated and $|v_i|  \ge n/b + 2 + 18\log n$.
\item\label{partinstshort} $v_i$ is partially instantiated and $|v_i|  < n/b + 2 + 18\log n$.
\item\label{noninst} $v_i$ has not been instantiated.
\end{enumerate}
($O(1)$ bits).
\item For every (partially) instantiated variable with $|v_i| <  n/b + 2 + 18\log n$ its current
value when the center has just been matched ($(b-1)(n/b + 2 + 18\log n)$ bits).
\end{itemize}
For recovering the string $w$ try to determine a matching of $\alpha$ and every 
$w_b\#\cdots w_2\#w_1$ with $w_i\in \{ 0, 1\}^{n/b}$ by backtracking. 
The matching is based on the recorded
values of the variables, where partially instantiated variables are extended until
the corresponding closing parenthesis is encountered. The matching starts at the recorded
occurrence of $\#$ in $\alpha$. If a variable $v$ is encountered in $\alpha$ and 
$v$ is of type \ref{instlong}, \ref{partinstlong},  \ref{partinstshort} or \ref{noninst} 
and in the latter two cases the current matching
extends the value to a length at least $n/b + 2 + 18\log n$, then the matching fails 
due to the Non-Matching Lemma.
If a variable occurs that has not been instantiated, the matching fails as well.
If for $w_b\#\cdots w_2\#w_1$ a matching has been determined, the string 
$w  = w_1w_2\cdots w_b$ has been reconstructed.

The description of $w$ has length $(b-1)n/b + O(\log n) = n - n/b + O(\log n) $ leading to
a compression for $n$ sufficiently large and thus  contradicting the choice of $w$. \qed

\section{Decidability Results}

\begin{theorem}\label{matchingSLAPREnostar}
The Matching-Problem for PRE without closure operators over a single-letter
alphabet is log-complete for NP. The same statement holds for membership. 
\end{theorem}
{\bf Proof.} The upper bound is shown as in the proof of Theorem~6.2 in \cite{Aho90},
with the additional possibility of an empty variable.

For hardness we reduce the well-known NP-complete problem 3SAT to the Matching-Problem. 
Let $F$ be a boolean formula in CNF with $n$ variables and $m$ clauses. We 
introduce a subexpression $u_i = ((0)|(0))$ for every variable
$x_i$ with $ 1 \le i \le n$ and a subexpression $v_j = (\backslash k_j^1|\backslash k_j^2|\backslash k_j^3)$ 
for clause $c_j = (y_j^1 \vee  y_j^2 \vee y_j^3)$ with 
$$
k_j^p =
\begin{array}{cl}
3i-1  & \mbox{ if $y_j^p = x_i$,}\\
3i  & \mbox{ if $y_j^p = \neg x_i$.}
\end{array}
$$
Expression $\alpha$ is defined as $\alpha = u_1\cdots u_nv_1\cdots v_m$. Finally we let $s = 0^{n+m}$.

Suppose $F$ has a satisfying assignment. A matching can be obtained by choosing the first subexpression
of $((0)|(0))$ for every boolean variable assigned the value `true' and the second subexpression otherwise.
The $v_1,\ldots, v_m$ are matched by choosing a satisfied literal for every clause, to which an instantiated
variable of the PRE corresponds. Conversely a matching can be otained only if at least one variable of the PRE
is instantiated for every $v_j$, which induces a partial assignment to the boolean variables. Variables not involved
in the matching can be set to an arbitrary value.

Notice that by construction $\alpha$ generates at most the string $s = 0^{n+m}$, such that the reduction works
for membership as well. \qed

Greibach's Theorem \cite{Greibach68} states that any nontrivial property $P$ of a class $C$ of formal languages 
over an alphabet $\Sigma\cup\{ \# \}$ is undecidable, provided that the following conditions
are satisfied:
\begin{enumerate}
\item\label{finitedescr} The languages in $C$ have finite descriptions.
\item\label{regular} $C$ contains every regular language over $\Sigma\cup\{ \# \}$.
\item\label{closure} For descriptions of languages $L_1, L_2 \in C$ and regular language $R\in C$, descriptions of  $L_1R$, $RL_1$, and $L_1\cup L_2$ can be computed effectively.
\item\label{universality} Universality ($L = \Sigma^*$?) is undecidable for $L \in C$ with $L \subseteq \Sigma^*$.
\item\label{cquotient} $P$ is closed under quotient by each symbol in $\Sigma\cup\{ \# \}$.
\end{enumerate}
Let $C$ be the class of languages characterized by PRE. The finite desciptions are PRE 
(Property~\ref{finitedescr}), regular languages are characterized by PRE without
variables (Property~\ref{regular}), PRE can be composed (Property~\ref{closure}) and
Freydenberger \cite{Freydenberger11} has shown universality to be undecidable 
(Property~\ref{universality}). For quotient (Property~\ref{cquotient}) 
we make use of the closure of PRE under intersection
with regular sets \cite{CS09}. After forming the intersection with $a(\Sigma\cup\{ \# \})^*$ for 
a symbol $a$, for every variable $k$ including the initial $a$ each occurrence 
$\backslash k$ is replaced by $a\backslash k$ and then the initial $a$ is removed.
Freydenberger's results of undecidability of 
regularity and cofiniteness follow from Greibach's Theorem. We can extend this list by
context-freeness.

\begin{theorem}\label{equivalencePREnostar}
Inequivalence of PRE without closure operators over a single letter alphabet is log-complete for
$\Sigma^{P}_2$.
\end{theorem}
{\bf Proof.} We notice that due to the lack of the closure operator the sets characterized
by the PRE of the considered form are finite and an occurrence of a variable can at 
most double the length of the longest
string characterized by an expression. Membership of the inequivalence problem 
in $\Sigma^{P}_2$ can be shown with an NP-machine $M$ that has 
access to an oracle that for a PRE $\beta$ and a string $0^n$ (where $n$ is encoded in binary)
solves the membership problem. Let $\alpha_1, \alpha_2$ be the PRE in the input of $M$.
Machine $M$ guesses a string $0^n$ in the symmetric difference of the languages defined 
by $\alpha_1$ and $\alpha_2$, where $n \le 2^{\max(|\alpha_1|, |\alpha_2|)}$, and 
in turn passes 
$n$ in binary and $\alpha_1$ resp.\ $\alpha_2$ to the oracle. The input is accepted if
exactly one of the answers returned by the oracle is positive.

For $\Sigma^{P}_2$-hardness we describe a log-space reduction from the inequivalence problem
for integer expressions denoted by N-INEQ to the inequivalence of PRE. It is known that
N-INEQ is log-space complete for $\Sigma^{P}_2$ \cite[Theorem~5.2]{SM73}. 
Integer expressions are expressions defining sets of nonnegative integers. 
The binary operations are $+$ (pairwise addition) and
$\cup$ (union), a singleton set is defined by the corresponding 
integer in binary notation without leading zeroes.
The problem N-INEQ is defined as follows: Given
two integer expressions $\gamma_1$,  $\gamma_2$, determine whether they define different sets of numbers. 
The reduction will transfom an integer expression $\gamma$ into a PRE $\alpha$ such that
$$ n \mbox{ is in the set defined by } \gamma \Leftrightarrow 0^n \in L(\alpha).$$
Addition of  integers corresponds to concatenation of strings and union is an operation available
in PRE. It remains to describe how to concisely encode a string $0^n$. The encoding of
1 is the string 0. Suppose $u$ is the encoding of $\lfloor n/2 \rfloor$. If $n$ is even, then
$(u')\backslash 1$ encodes $0^n$, if $n$ is even, then $(u')\backslash 10$ encodes $0^n$, 
where $u$ is $u$ with all references incremented by one (because of the additional pair
of parentheses). Clearly the length of the encoding is $O(\log n)$. \qed

The upper bound of the previous theorem can be extended to PRE over larger alphabets as follows.
The separating string is communicated to the oracle in compressed form as a
straight-line program (SLP). An SLP is a context-free grammar that generates
exactly one string. Such an SLP can be generated from a PRE starting with the 
innermost parentheses. Each definition of a variable is nondeterminstically converted into
a context-free production with a fresh nonterminal symbol that generates a string described
by the subexpression in parentheses. Then each orrurrence of the variable is replaced with 
the non-terminal symbol. When all variables have been replaced, the alternatives are eliminated
and the resulting string is taken as the right-hand side of the initial nonterminal. 
The oracle uses the same method for encoding a word described by the PRE in its input.
Several methods for comparing SLP-compressed strings are known, see \cite[Section~5]{Lohrey12}.
\begin{proposition}
Inequivalence of PRE without closure operators is in $\Sigma^{P}_2$.
\end{proposition}

For general PRE over a single letter alphabet we do not have a positive or negative decidability 
result for equivalence, but we
can provide evidence that a decision procedure will not be obvious by linking it to Fermat Primes 
(primes of the form $2^{2^n}+1$) \cite[Sequence~A019434]{OEIS}.

\begin{theorem}\label{equivalencePREFermat}
If for PRE over a single letter alphabet equivalence is effectively decidable, we can solve the 
open problem whether 3, 5, 17, 257 and 65537 are the only prime Fermat numbers.
\end{theorem}
{\bf Proof.} 
The PRE $\alpha_1$ describes all strings of length greater than two
except those with a length of the form $2^n+1$ for $n \ge 1$: 
$$\alpha_1 = ((aa)+a)\backslash 1*a.$$
Expression $\alpha_2$ describes all nonempty strings having a length with
a proper divisor, thus omitting all primes and one:
$$\alpha_2 = (a+a)\backslash 1+.$$
We combine the expressions $\alpha_1$ and $\alpha_2$ with strings of length
3, 5, 17, 257 and 65537 (the known Fermat primes) and add 0, 1, and 2:
$$\alpha_3 = ((aa)+a)\backslash 1*a|(a+a)\backslash 3+|a^0|a^1|a^2|a^3|a^5|a^{17}|a^{257}|a^{65537}.$$
Now $\alpha_3$ is equivalent to $a*$ if and only if no additional Fermat prime
exists. \qed

\section{Discussion}
We have established connections between PRE and classical models of computation within NSPACE($\log n$).
The hierarchy of PRE with a growing number of variables has been strengthened to expressions without 
nested variables and a fixed alphabet.
The NP-hardness result for the Matching-Problem has been strengthened to a single letter alphabet
making use of uninstantiated variables. It is open whether a similar result is possible under the syntactic
restrictions of \cite{Larsen98}. The complexity of equivalence of PRE without star over a single letter alphabet
has been characteriized and we cojecture  that equivalence for general PRE over a single letter alphabet is undecidable.

\section*{Acknowledgement}
I am very grateful to Jeff Shallit for a correction concerning an earlier version of this paper.

\end{document}